\documentclass[twocolumn,prd,aps,showpacs,showkeys,amsmath,amssymb]{revtex4-1}
\usepackage{bm}

\usepackage{color}
\usepackage{amsmath}
\usepackage{amsfonts}
\usepackage{verbatim}
\usepackage{amssymb}
\usepackage{graphicx}
\usepackage{epstopdf}
\usepackage{mathrsfs}
\usepackage{epsfig}
\usepackage{slashed}
\usepackage{bbold}

\providecommand{\U}[1]{\protect\rule{.1in}{.1in}}
\newcommand{\bl}{\boldsymbol}

\newcommand{\ph}{\phantom}

\newcommand{\eq}{\,=\,}
\newcommand{\ma}{\,+\,}
\newcommand{\me}{\,-\,}

\newcommand{\arctanh}{\textrm{arctanh}}

\begin{document}

\title{Integrability of the Dirac Equation on Backgrounds that are the Direct Product of Bidimensional Spaces}
\author{Jo\'{a}s Ven\^{a}ncio and Carlos Batista}
\email[]{joasvenancio@df.ufpe.br, carlosbatistas@df.ufpe.br}
\affiliation{Departamento de F\'{\i}sica, Universidade Federal de Pernambuco,
Recife, Pernambuco  50740-560, Brazil}


\begin{abstract}
The field equation for a spin $1/2$ massive charged particle propagating in spacetimes that are the direct product of 2-dimensional spaces is separated. Moreover, we use this result to attain the separability of the Dirac equation in some specific static black hole solutions whose horizons have topology $\mathbb{R}\times  \mathbb{S}^2 \times \cdots \times \mathbb{S}^2$.
\end{abstract}
\maketitle

\section{Introduction}

Besides the detection of gravitational radiation and observation of the direct interaction between objects via gravitation, the most natural and simple way to probe the gravitational field permeating our spacetime is by letting other fields interact with it. This is the main reason why the study of scalar fields, spin $1/2$ fields and gauge fields (abelian and non-abelian) propagating in curved spacetimes plays a central role on the study of general relativity and any other theory of gravity. Moreover, through the investigation of these interactions one can test the stability of certain gravitational configurations, such as black holes for example. Nevertheless, in a general spacetime it is quite difficult to integrate and even separate the equation of motion for these fields. Luckily, some of the most important spacetimes are endowed with geometrical structures that allow the integration of these field equations. For instance, this is the case for the Schwarzschild spacetime, that possesses four Killing vector fields, and the Kerr spacetime, which has two Killing vectors and one Killing-Yano tensor \cite{Carter-constant,Walk-Pen}. Indeed, in Refs. \cite{Price:1971fb,Price:1972pw} the separation and asymptotic integration of some of the mentioned field equations have been attained in the Schwarzschild background, while in Refs. \cite{Teukolsky:1972my,Unruh:1973bda,Chandrasekhar:1976ap,Page:1976jj} the problem is tackled in the Kerr spacetime. Many of these results are summarized by the celebrated Teukolsky master equation, which is the radial part of the equation of motion for a field of generic spin propagating in the four-dimensional Kerr background \cite{Teukolsky:1973ha,Press:1973zz,Teukolsky:1974yv}.

In the past forty years, a great amount of research has been headed toward the investigation of higher-dimensional spacetimes, specially due to their  interest for String theory \cite{Mukhi:2011zz}, which requires the spacetime to have $10$ dimensions, and because of applications in field theory by means of the AdS/CFT correspondence \cite{Maldacena:1997re}. Moreover, there are many other theories that tries to explain our current understanding of the Universe by means higher-dimensional theories \cite{Csaki:2004ay}. Aiming applications on some of these fields, in the present article we shall workout the separation of the Dirac equation for a massive spin $1/2$ charged particle propagating in the black hole background described in Ref. \cite{Batista-BH}, which is a static black hole whose horizon topology is $\mathbb{R}\times  \mathbb{S}^2 \times \cdots \times \mathbb{S}^2$. One interesting feature of this black hole is that, in addition to the electric charge, it has a magnetic charge, differently from the higher-dimensional generalization of the Reissner-Nordstrom solution \cite{Tangherlini}, which only has electric charge. Thus, in spite of the static character of the black hole considered here, the physics involved can be quite rich.

One of the applications of the calculations performed here is on the study of the quasi-normal modes of the Dirac field. The notion of quasi-normal modes and their spectrum are of great physical relevance, inasmuch as these are the modes that survive for a longer time when a background is perturbed and, therefore, these are the configurations that are generally measured by experiments \cite{Kokkotas:1999bd,Nollert:1999ji,Berti:2009kk}. Therefore, this theme acquired even greater importance after the recent measurement of gravitational radiation \cite{Abbott:2016blz}. In addition, the quasi-normal modes are of great relevance for studying the stability of certain solutions \cite{Zhidenko:2009zx}. Another interesting application of the results presented in the sequel is on the investigation of superradiance phenomena for the spin $1/2$ field. Although bosonic fields like scalar, electromagnetic and gravitational fields can exhibit superradiant behaviour in four-dimensional Kerr spacetime \cite{Rosa:2016bli}, curiously, this is not the case for the Dirac field \cite{Gueven:1977dq}. Thus, it would be interesting to investigate whether an analogous thing happens in the background considered here.

The outline of the paper is the following. Section \ref{Sec.Notation} sets the notation and the conventions used throughout the article. Then, in Sec. \ref{Sec.Conformal} we work out how the Dirac operator behaves under a conformal transformation of the metric, a result that will be of great value for the separation of the Dirac equation in the black hole background. In Sec. \ref{SecSeparability} we show that a generalized Dirac equation can be separated in a space that is the direct product of 2-dimensional spaces. Sec. \ref{Sec.BHs} uses the results obtained in the previous sections to attain the separation of the Dirac equation in a black hole background. Finally, in Sec. \ref{Sec.Conclusions} we sum up what have been done and digress about future applications of this research.

\section{Notation and Conventions}\label{Sec.Notation}


In what follows, we shall deal with an even-dimensional manifold $M$ endowed with a metric $\bl{g}$ and a torsion-free connection $\nabla$ that is compatible with the metric. Regarding the signature, in general, we shall not restrict ourselves to a specific choice of signature, although our main motivation are the Lorentzian spacetimes. The dimension will be denoted by $d=2n$. Our indices conventions are: the Greek letters from the middle of the alphabet ($\mu,\,\nu$) are coordinate indices and range from $1$ to $2n$; the Greek letters from the beginning of the alphabet ($\alpha,\,\beta,\,\varepsilon$) run from $1$ to $2n$ and label the vector fields of an orthonormal frame $\{\bl{e}_\alpha\}$; lowercase Latin indices with and without tildes ($a,\,b,\,\tilde{a},\,\tilde{b}$) range from $1$ to $n$ and are also used to label the vector fields of an orthonormal frame, but in a pairwise form $\{\bl{e}_a,\bl{e}_{\tilde{a}}\}$, which will be quite suitable to our intent, as will be clear in the sequel; the indices $(\ell,\,\tilde{\ell})$ run from $2$ to $n$ and serve to label the angular directions of the black hole spacetime considered here; finally, the indices $(s,\,s_1,\,s_2,\,\cdots)$ can take the values $\pm1$ and label spinorial degrees of freedom. In what follows, Einstein's summation convention is adopted for pairs of equal indices that are in opposite places, one up and the other down. But equal indices that are in the same position, both up or both down, should not be summed in principle, unless an explicit symbol of sum is included.

By an orthonormal frame we mean that
\begin{equation*}
\bl{g}(\bl{e}_\alpha,\bl{e}_\beta)\eq \delta_{\alpha\beta}  \quad \leftrightarrow \quad  \left\{
                                                                        \begin{array}{ll}
                                                                          \bl{g}(\bl{e}_a,\bl{e}_b)\eq \delta_{ab} \\
                                                                       \bl{g}(\bl{e}_a,\bl{e}_{\tilde{b}})\eq 0 \\
                                                                          \bl{g}(\bl{e}_{\tilde{a}},\bl{e}_{\tilde{b}})\eq \delta_{\tilde{a}\tilde{b}}
                                                                        \end{array}
                                                                      \right. \,,
\end{equation*}
where it has been used the fact that the indices $a$ and $b$ can be thought as labeling the first $n$ vector fields of the orthonormal frame $\{\bl{e}_\alpha\}$, while $\tilde{a}$ and $\tilde{b}$ label the remaining $n$ vectors of the frame $\{\bl{e}_\alpha\}$. If the signature is not Euclidean some of the vector fields $\bl{e}_\alpha$ might be imaginary in order for the frame be orthonormal. The derivatives of the frame vector fields determine the spin connection according to the following relation
\begin{equation*}
 \nabla_{\alpha} \bl{e}_\beta \eq \omega_{\alpha\beta}^{\ph{\alpha\beta}\varepsilon}\,\bl{e}_\varepsilon  \,.
\end{equation*}
Indices of the spin connection are raised and lowered with $\delta^{\alpha\beta}$ and $\delta_{\alpha\beta}$ respectively, so that frame indices can be raised and lowered unpunished. Since the metric is covariantly constant, it follows that $\omega_{\alpha\beta\varepsilon} \eq \omega_{\alpha[\beta\varepsilon]}$, where indices inside square brackets are anti-symmetrized. Analogously, indices enclosed by round brackets are assumed to be symmetrized. The Dirac matrices  $\gamma_\alpha$ are $2^n\times 2^n$ matrices obeying the Clifford algebra,
\begin{equation}\label{CliffordAlgebra}
 \gamma_{(\alpha}\gamma_{\beta)}\eq  \frac{1}{2} \left( \gamma_\alpha\,\gamma_\beta \ma  \gamma_\beta\,\gamma_\alpha \right) \eq 
 \delta_{\alpha\beta}\,   \mathbb{1}    \, ,
\end{equation}
with $ \mathbb{1}$ standing for the $2^n\times 2^n$ identity matrix. The covariant derivative of a spinorial field $\bl{\psi}$ is, then, given by
\begin{equation*}
  \nabla_\alpha \bl{\psi} \eq
\partial_\alpha \bl{\psi} \me \frac{1}{4}\,\omega_{\alpha}^{\ph{\alpha}\beta\varepsilon}\,\gamma_\beta \gamma_\varepsilon \bl{\psi} \,,
\end{equation*}
with $\partial_\alpha$ denoting the partial derivative along the vector field $\bl{e}_\alpha$. Given an orthonormal frame $\{\bl{e}_\alpha\}$, we can define the dual frame of 1-forms $\{\bl{E}^\alpha\}$, which is defined to be such that
\begin{equation*}
  \bl{E}^\alpha (\bl{e}_\beta) \eq \delta^\alpha_{\;\beta} \,.
\end{equation*}
Thus, if $\{x^\mu\}$ is a local coordinate system in our manifold $M$, the line element can be written as
\begin{equation*}
  ds^2 \eq g_{\mu\nu}dx^\mu dx^\nu \eq \delta_{\alpha\beta}\bl{E}^\alpha \bl{E}^\beta \eq \bl{E}^\alpha \bl{E}_\alpha\,,
\end{equation*}
where $g_{\mu\nu}\eq \bl{g}(\partial_\mu,\partial_\nu)$ are the components of the metric in the coordinate frame.

\section{Conformal Transformation and the Dirac Operator}\label{Sec.Conformal}

In this section we shall obtain the conformal transformation of the Dirac operator, which will be of future relevance in order to simplify the Dirac equation in the spacetime of our interest.

Let $\hat{\bl{g}}$ be a metric that is conformally related to our initial metric, $\hat{\bl{g}} = \Omega^2 \bl{g}$, with $\Omega$ being a positive definite function throughout the manifold. Then, if $\{\hat{\bl{e}}_\alpha\}$ is an orthonormal frame with respect to the metric $\hat{\bl{g}}$ we have
\begin{equation}\label{Ee-conformal}
   \hat{\bl{E}}^\alpha \eq \Omega\, \bl{E}^\alpha \;\; \textrm{and} \;\; \hat{\bl{e}}_\alpha \eq \Omega^{-1}\, \bl{e}_\alpha\,.
\end{equation}
Since we are dealing with metric-compatible connections, a change of metric lead to a different spin connection, which is defined by the following relation
\begin{equation}\label{nabla-conformal}
 \hat{\nabla}_{\alpha} \hat{\bl{e}}_\beta \eq \hat{\omega}_{\alpha\beta}^{\ph{\alpha\beta}\varepsilon}\,\hat{\bl{e}}_\varepsilon  \,.
\end{equation}
Using the identity $\hat{g}_{\mu\nu} = \Omega^2 g_{\mu\nu}$ to find the relation between the Levi-Civita symbols of the two metrics and using Eqs. (\ref{Ee-conformal}) and (\ref{nabla-conformal}), we eventually arrive at the following expression:
\begin{equation}\label{omega-conformal}
  \hat{\omega}_{\alpha}^{\ph{\alpha}\beta\varepsilon} \eq
 \Omega^{-1}\,\omega_{\alpha}^{\ph{\alpha}\beta\varepsilon}  \ma 2 \, \Omega^{-2}\, \partial^{[\beta}\Omega \,\, \delta^{\varepsilon]}_{\;\;\alpha}\,.
\end{equation}

Now, let us obtain how the Dirac operator, defined by $D = \gamma^\alpha \nabla_\alpha$, behaves under conformal transformations. If $\bl{\psi}$ is a spinorial field, let us define
\begin{equation*}
  \hat{\bl{\psi}} \eq  \Omega^p \, \bl{\psi}\,,
\end{equation*}
with $p$ being a constant parameter that will be conveniently chosen in the sequel. Then, using Eq. (\ref{omega-conformal}) we eventually obtain the identity below:
\begin{align*}
  \hat{D}  \hat{\bl{\psi}} & \eq \gamma^\alpha\, \hat{\nabla}_\alpha \left( \Omega^p \, \bl{\psi} \right)  \\
& \eq \Omega^{p-1} D\bl{\psi} \ma \left(p+n-\frac{1}{2}\right) \Omega^{p-2} (\partial_\alpha\Omega) \gamma^\alpha \bl{\psi} \,.
\end{align*}
Thus, choosing $p$ to be $\frac{1}{2}-n$, it follows that
\begin{equation}\label{DiracOp-conformal}
 D \bl{\psi} \eq \Omega^{(n+\frac{1}{2})}\, \hat{D}  \hat{\bl{\psi}}  \,,\; \textrm{where} \;\, \hat{\bl{\psi}} \eq  \Omega^{(\frac{1}{2}-n)} \, \bl{\psi} \,.
\end{equation}
In particular, this relation enables to investigate the conformal invariance of the Dirac equation. Indeed, if $\bl{\psi}$ is a spinorial field of mass $m$ that obeys Dirac equation in the spacetime with metric $\bl{g}$ then
\begin{equation}\label{DiracEq-conformal}
  D \bl{\psi} \eq m\,\bl{\psi} \; \Rightarrow\; \hat{D} \hat{\bl{\psi}} \eq (\Omega^{-1}\,m)\,\hat{\bl{\psi}}\,.
\end{equation}
Since, generally, $\Omega$ is a non-constant function, it follows that the massive Dirac equation is not conformally invariant, whereas the massless Dirac equation is invariant under conformal transformations. In spite of the lack of conformal invariance of the Dirac equation with mass, Eq. (\ref{DiracEq-conformal}) will be of great help for the separation of the Dirac equation in black hole the spacetime considered in this work.

\section{Direct Product Spaces and the Separability of the Dirac Equation} \label{SecSeparability}

The goal of the present section is to show that the Dirac equation minimally coupled to an electromagnetic field is separable in spaces that are the direct product of bidimensional spaces.

Let $(M,\hat{\bl{g}})$ be a $2n$-dimensional space that is the direct product of $n$ bidimensional spaces, namely the space can be covered by coordinates
$\{x^1,y^1,x^2,y^2,\cdots,x^n,y^n\}$  such that the line element is given by
\begin{equation}\label{MetricConf}
  d\hat{s}^2 \eq \sum_{a=1}^{n}\, d\hat{s}_a^2 \eq \sum_{a=1}^{n}\, ( \hat{\bl{E}}^a\hat{\bl{E}}^a \ma \hat{\bl{E}}^{\tilde{a}} \hat{\bl{E}}^{\tilde{a}} ) \,,
\end{equation}
where the 2-dimensional line elements $d\hat{s}_a^2$ and the 1-forms $\hat{\bl{E}}^{a}$ and  $\hat{\bl{E}}^{\tilde{a}}$ depend just on the two coordinates corresponding to their bidimensional spaces. For instance, $d\hat{s}_1^2$, $\hat{\bl{E}}^{1}$ and  $\hat{\bl{E}}^{\tilde{1}}$ depend just on the differentials $dx^1$ and $dy^1$ and theirs components depend just on the coordinates $x^1$ and $y^1$. In such a case, the only components of the spin connection that can be non-vanishing are
\begin{equation}\label{Spinconnetion}
  \hat{\omega}_{aa\tilde{a}} = -\, \hat{\omega}_{a\tilde{a}a} \quad  \textrm{and} \quad
  \hat{\omega}_{\tilde{a}a\tilde{a}} = -\, \hat{\omega}_{\tilde{a}\tilde{a}a} \,.
\end{equation}
Thus, for example, $\hat{\omega}_{aa\tilde{b}} = 0$  and $\hat{\omega}_{aab} = 0$ whenever $a\neq b$. Furthermore, the non-zero components of the spin connection  associated to the index $a$ depend just on the coordinates $x^a$ and $y^a$. For instance, $\hat{\omega}_{11\tilde{1}}$ is a function that depends just on $x^1$ and $y^1$.

In order to accomplish the separability of the Dirac equation, it is necessary to use a suitable representation for the Dirac matrices. In what follows, the $2\times2$ identity matrix will be denoted by $\mathbb{I}$, while the usual notation for the Pauli matrices is going to be adopted:
\begin{equation*}
\sigma_1 \eq \left[
                  \begin{array}{cc}
                    0 & 1 \\
                    1 & 0 \\
                  \end{array}
                \right] \, , \;\;
                \sigma_2 \eq \left[
                  \begin{array}{cc}
                    0 & -i \\
                    i & 0 \\
                  \end{array}
                \right]  \, , \;\;
                \sigma_3 \eq \left[
                  \begin{array}{cc}
                    1 & 0 \\
                    0 & -1 \\
                  \end{array}
                \right]\,.
\end{equation*}
Using this notation, a convenient representation of the Dirac matrices in $2n$ dimensions is the following:
\begin{align}
  \gamma_{a} & \eq \underbrace{\sigma_3\otimes \cdots \otimes  \sigma_3}_{(a-1)\;\textrm{times}}  \otimes  \sigma_1  \otimes
  \underbrace{\mathbb{I} \otimes \cdots \otimes  \mathbb{I}}_{(n-a)\;\textrm{times}} \,,\nonumber \\
  \quad \label{DiracMatrices}\\
  \gamma_{\tilde{a}} & \eq \underbrace{\sigma_3\otimes \cdots \otimes  \sigma_3}_{(a-1)\;\textrm{times}}  \otimes  \sigma_2  \otimes
  \underbrace{\mathbb{I} \otimes \cdots \otimes  \mathbb{I}}_{(n-a)\;\textrm{times}} \,.\nonumber
\end{align}
Indeed, we can easily check that the Clifford algebra given in Eq. (\ref{CliffordAlgebra}) is properly satisfied by the above matrices. Regarding the spinors, it is useful to define the following column vectors:
\begin{equation}\label{SpinBasis1}
  \xi^+ \eq \left[
              \begin{array}{c}
                1 \\
                0 \\
              \end{array}
            \right] \quad \textrm{and} \quad
             \xi^- \eq \left[
              \begin{array}{c}
                0 \\
                1 \\
              \end{array}
            \right] \,.
\end{equation}
If we assume that the index $s$ can take the values ``$+1$'' and  ``$-1$'', the action of the Pauli matrices on the above column vectors can be summarized quite concisely as
\begin{equation}\label{PauliAction}
  \sigma_1\xi^s \eq \xi^{-s} \;,\;\; \sigma_2\xi^s \eq i\,s\,\xi^{-s}  \;,\;\;  \sigma_3\xi^s \eq s\,\xi^{s} \,.
\end{equation}
The spinor space, in which the Dirac matrices act, can be spanned by the direct product of the elements $\xi^{s}$ $n$ times. More precisely, a general spinor field can be written as
\begin{equation}\label{SpinBasis2}
  \hat{\bl{\psi}} \eq \sum_{\{s\}} \hat{\psi}^{s_1s_2\cdots s_n} \; \xi^{s_1}\otimes  \xi^{s_2}\otimes  \cdots \otimes  \xi^{s_n} \,,
\end{equation}
in which $\hat{\psi}^{s_1s_2\cdots s_n}$ stands for the components of the spinor and the sum over $\{s\}$ means the sum over all possible values of the set $\{s_1,s_2,\cdots,s_n\}$. Since, every $s_a$ can take two values, it follows that this sum comprises $2^n$ terms, which is the number of components of a spinor in $d=2n$ dimensions. Using this basis, we can easily compute the action of the Dirac matrices on the spinor field. Indeed, using Eqs.  (\ref{DiracMatrices}), (\ref{PauliAction}) and (\ref{SpinBasis2}) we have:
\begin{widetext}
\begin{align*}
 \gamma_a \hat{\bl{\psi}} & \eq \sum_{\{s\}}(s_1 s_2 \cdots s_{a-1}) \hat{\psi}^{s_1s_2\cdots s_n} \;
\xi^{s_1}\otimes \xi^{s_2}\otimes \cdots \otimes \xi^{s_{a-1}}\otimes  \xi^{-s_{a}} \otimes \xi^{s_{a+1}} \otimes  \cdots \otimes  \xi^{s_n} \\
& \eq  \sum_{\{s\}}(s_1 s_2 \cdots s_{a})\,s_a\, \hat{\psi}^{s_1s_2\cdots s_{a-1} (-s_a) s_{a+1}\cdots s_n} \;
\xi^{s_1}\otimes \xi^{s_2}\otimes \cdots \otimes \xi^{s_{a-1}}\otimes  \xi^{s_{a}} \otimes \xi^{s_{a+1}} \otimes  \cdots \otimes  \xi^{s_n} \,.
\end{align*}
Where from the first to the second line we have changed the index $s_a$ to $-s_a$, which does not change the final result, since we are summing over all values of $s_a$, which comprise the same list of the values of $-s_a$. Moreover, we have used that $(s_a)^2=1$. Analogously, we have:
\begin{align*}
 \gamma_{\tilde{a}} \hat{\bl{\psi}} & \eq \sum_{\{s\}}(s_1 s_2 \cdots s_{a-1})(i \, s_a)\, \hat{\psi}^{s_1s_2\cdots s_n} \;
\xi^{s_1}\otimes \xi^{s_2}\otimes \cdots \otimes \xi^{s_{a-1}}\otimes  \xi^{-s_{a}} \otimes \xi^{s_{a+1}} \otimes  \cdots \otimes  \xi^{s_n} \\
& \eq  -i\sum_{\{s\}}(s_1 s_2 \cdots s_a)\, \hat{\psi}^{s_1s_2\cdots s_{a-1} (-s_a) s_{a+1}\cdots s_n} \;
\xi^{s_1}\otimes \xi^{s_2}\otimes \cdots \otimes \xi^{s_{a-1}}\otimes  \xi^{s_{a}} \otimes \xi^{s_{a+1}} \otimes  \cdots \otimes  \xi^{s_n}
\end{align*}

Now, we have the tools to try to separate the general equation
\begin{equation}\label{Dif.EQ1}
  \left[\hat{D} \me (\hat{A}_a \gamma^a + \hat{A}_{\tilde{a}} \gamma^{\tilde{a}}) \right] \hat{\bl{\psi}} \eq \hat{m} \, \hat{\bl{\psi}}
\end{equation}
in its 2-dimensional blocks, where $\hat{D}$ is the Dirac operator of the space with metric (\ref{MetricConf}) while  $\hat{A}_a$, $\hat{A}_{\tilde{a}}$ and $\hat{m}$ are arbitrary functions. In order to accomplish this goal, we shall assume that the components of the spinor field (\ref{SpinBasis2}) take the separable form
\begin{equation}\label{SpinorSeparable}
  \hat{\psi}^{s_1s_2\cdots s_n} \eq \hat{\psi}_1^{s_1}(x^1,y^1)\,\hat{\psi}_2^{s_2}(x^2,y^2)\,\cdots \hat{\psi}_n^{s_n}(x^n,y^n) \,.
\end{equation}
Using this hypothesis and noting that the Dirac operator is $\hat{D} = \gamma^a \hat{\nabla}_a \ma  \gamma^{\tilde{a}}\hat{\nabla}_{\tilde{a}}$, it follows that Eq. (\ref{Dif.EQ1}) is given by:
\begin{multline*}
\sum_{a=1}^{n}  \sum_{\{s\}} (s_1 s_2 \cdots s_{a}) \, \hat{\psi}_1^{s_1}\,\cdots\,\hat{\psi}_{a-1}^{s_{a-1}}
\hat{\psi}_{a+1}^{s_{a+1}} \,\cdots\, \hat{\psi}_n^{s_n}  \\
  \left[ i\,s_a\,\left( \hat{\partial}_a + \frac{1}{2} \hat{\omega}_{\tilde{a}a\tilde{a}} - \hat{A}_{a} \right) +  \left( \hat{\partial}_{\tilde{a}} + \frac{1}{2} \hat{\omega}_{a\tilde{a}a} - \hat{A}_{\tilde{a}} \right) \right]
  \hat{\psi}_{a}^{(-s_{a})} \, \xi^{s_1}\otimes \xi^{s_2}\otimes \cdots \otimes \xi^{s_{n}} \eq \\
  i\,\hat{m} \, \sum_{\{s\}}  \hat{\psi}_1^{s_1}  \hat{\psi}_2^{s_2}\,\cdots\,\hat{\psi}_{n}^{s_{n}} \xi^{s_1}\otimes \xi^{s_2}\otimes \cdots \otimes \xi^{s_{n}} \,,
\end{multline*}
where by $\hat{\partial}_a$  and $\hat{\partial}_{\tilde{a}}$ we mean the derivatives along the vector fields $\hat{\bl{e}}_a$ and  $\hat{\bl{e}}_{\tilde{a}}$, namely $(\hat{e}_a)^{\;\mu}\partial_\mu$ and $(\hat{e}_{\tilde{a}})^{\;\mu}\partial_\mu$ respectively. In order for the latter equation to be separable in blocks depending only on the coordinates $\{x^1,\,y^1\}$,  $\{x^2,\,y^2\}$ and so on, the functions $\hat{A}^a$ and $\hat{A}^{\tilde{a}}$ must depend only on the two coordinates $\{x^a,\,y^a\}$ and the function $\hat{m}$ must be a sum of functions depending on these pairs of coordinates:
\begin{equation}\label{Am}
  \hat{A}_a \eq \hat{A}_a(x^a,y^a) \;, \quad \hat{A}_{\tilde{a}} \eq \hat{A}_{\tilde{a}}(x^a,y^a)  \;, \quad
  \hat{m} \eq \sum_{a=1}^{n} \hat{m}_a(x^a,y^a) \,.
\end{equation}
With these necessary assumptions for attaining separability, we are left with the following equation:
\begin{equation}\label{Dif.EQ2}
 \sum_{a=1}^{n} \, \left[\, (s_1 s_2 \cdots s_{a})  \,  \frac{1}{\hat{\psi}_{a}^{s_{a}}} \,
 \slashed{D}_a^{s_a} \hat{\psi}_{a}^{(-s_{a})} - i\,\hat{m}_a \, \right] \eq 0 \,,
\end{equation}
where the operator $\slashed{D}_a^{s_a}$ used above is defined by
\begin{equation}\label{Dslash}
  \slashed{D}_a^{s_a} = i\,s_a\,\left(  \hat{\partial}_a + \frac{1}{2} \hat{\omega}_{\tilde{a}a\tilde{a}} - \hat{A}_{a} \right)
   +  \left( \hat{\partial}_{\tilde{a}} + \frac{1}{2} \hat{\omega}_{a\tilde{a}a} - \hat{A}_{\tilde{a}} \right) \,.
\end{equation}
\end{widetext}
Since each term in the sum over $a$ in Eq. (\ref{Dif.EQ2}) depends just on the two coordinates $\{x^a,\,y^a\}$, it follows that each of these terms in the sum must be a constant, otherwise they could not sum to zero. Let us denote these separation constants by $i\eta_a$. Thus, Eq. (\ref{Dif.EQ2}) requires that $\sum_a \eta_a = 0$. However, it is worth noting that Eq. (\ref{Dif.EQ2}) provides not only one equation but rather a total of $2^n$ independent equations, since for each choice of $\{s\} \equiv \{s_1,s_2,\cdots,s_n\}$ we have one equation. For each of these equations we can have different separation constants. Therefore, $\eta_a$ can depend on the choice  of $\{s\}$, so that it is appropriate to write these separation constants as $i\eta_a^{\{s\}}$. Then, we arrive at the following equations:
\begin{equation}\label{Dif.EQ3}
  (s_1 s_2 \cdots s_{a}) \slashed{D}_a^{s_a}  \hat{\psi}_{a}^{(-s_{a})} \eq i\,(\hat{m}_a \ma \eta_a^{\{s\}})  \hat{\psi}_{a}^{s_{a}} \,.
\end{equation}
These equations enable us to integrate the fields $\hat{\psi}_{a}^{s_{a}}$ and, therefore, find the solutions for the generalized Dirac equation (\ref{Dif.EQ1}). Although these equations are first order differential equations, they are coupled in pairs, namely the equations involving the field $\hat{\psi}_{a}^{+}$ have $\hat{\psi}_{a}^{-}$ as source and vice-versa. Therefore, after unraveling this system, we are left with a decoupled second order differential equation for each component $\hat{\psi}_{a}^{s_{a}}$, thus achieving the separability that we were looking for.

Note that, in accordance with Eq. (\ref{Dif.EQ2}), the separation constants must obey
\begin{equation}\label{SumLambda}
  \sum_{a=1}^{n}\,  \eta_a^{\{s\}} \eq 0 \,.
\end{equation}
Since the collective ``index'' $\{s\}$ can take $2^n$ values, the latter equation comprise $2^n$ constraints. Let us unravel these constraints.
Note that, since the left hand side of Eq. (\ref{Dif.EQ3}) is independent of $s_{a+1}$, $s_{a+2}$, $\cdots$, $s_n$, it follows that $\eta_a^{\{s\}}$ cannot depend on these indices. In particular, $\eta_1^{\{s\}}$ depends just on $s_1$, so that we can write
\begin{equation}\label{eta1-lambda1}
  \eta_1^{\{s\}} \eq s_1\,\kappa_1^{s_1} \,,
\end{equation}
where $\kappa_1^{s_1}$ is a pair of constants that depends just on $s_1$. Since $(s_a)^2 =1$, it follows that from Eq. (\ref{Dif.EQ3}) that we can write
\begin{equation}\label{Lambda1}
  \frac{1}{\hat{\psi}_{a}^{s_{a}}}\, \slashed{D}_a^{s_a}  \hat{\psi}_{a}^{(-s_{a})} \eq i\,(s_1 s_2 \cdots s_{a})  (\hat{m}_a \ma \eta_a^{\{s\}})   \,.
\end{equation}
Inasmuch as the left hand side of this equation depends just on $s_a$, it follows that the right hand side of this equation should, likewise, depend just on $s_a$. Therefore, we have
\begin{equation}\label{etaa}
  \hat{m}_a \ma \eta_a^{\{s\}} \eq (s_1 s_2 \cdots s_{a}) \, \kappa_a^{s_a} \quad \textrm{if} \quad a\geq 2 \,,
\end{equation}
where, for each $a$, $\kappa_a^{s_a}$ is a pair of parameters that depends just on $s_a$. Thus, for instance, $\kappa_a^{s_a}$ does not depend on $s_{a-1}$ and on $s_{a+1}$. But, since, in principle, $\hat{m}_a$ is non-constant, we cannot say that the parameters $\kappa_a^{s_a}$ are constant for $a\geq2$.   Nevertheless, taking the derivative of both sides of the latter equation, we have
\begin{equation*}
  \partial_\mu \hat{m}_a  \eq (s_1 s_2 \cdots s_{a}) \,\partial_\mu \kappa_a^{s_a}  \quad \textrm{if} \quad a\geq 2 \,.
\end{equation*}
Since the left hand side of the latter equation does not depend on $\{s\}$, it follows that $\partial_\mu \kappa_a^{s_a}$ must vanish, which, in its turn, implies that $\hat{m}_a$ should be constants for $a\geq 2$. But, if $\hat{m}_2(x^2,y^2)$, $\hat{m}_3(x^3,y^3)$, $\cdots$, $\hat{m}_n(x^n,y^n)$  are constants we can, without loss of generality, make all of them zero and absorb these constants in $\hat{m}_1(x^1,y^1)$. Therefore, we can say that a consistent separability process requires that
\begin{equation}\label{ha}
  \hat{m}_2 \eq \hat{m}_3 \eq \cdots \eq \hat{m}_n \eq 0 \,.
\end{equation}
Assuming this requirement to hold, Eqs. (\ref{eta1-lambda1}) and (\ref{etaa}) immediately lead to
\begin{equation}\label{eta-lambda}
  \eta_a^{\{s\}} \eq (s_1 s_2 \cdots s_{a}) \, \kappa_a^{s_a} \,,
\end{equation}
where, for each $a$, $\kappa_a^{s_a}$ is a pair of constants. Thus, the constraint (\ref{SumLambda}) now writes as
\begin{equation}\label{SumLambda2}
  \sum_{a=1}^{n}\, (s_1 s_2 \cdots s_{a}) \, \kappa_a^{s_a} \eq 0 \,.
\end{equation}
Note that this equation must hold for all possible choices of $\{s\}$. Now, let us manipulate this equation in order to solve such constraint.  Isolating  $\kappa_1^{s_1}$, we have
\begin{equation*}
  \kappa_1^{s_1} \eq - \, \sum_{a=2}^{n}\, (s_2 s_3 \cdots s_{a}) \, \kappa_a^{s_a} \,.
\end{equation*}
Note that the left hand side depends just on $s_1$. Therefore, the sum on right hand can depend just on $s_1$. However, since none of the terms in the sum depend on $s_1$ we conclude that this sum is a constant, namely:
\begin{equation*}
  \kappa_1^{s_1} \eq - c_1 \quad \textrm{and} \quad  \sum_{a=2}^{n}\, (s_2 s_3 \cdots s_{a}) \, \kappa_a^{s_a} \eq c_1 \,,
\end{equation*}
where $c_1$ is a constant that does not depend on $\{s\}$. Now, the latter equation can be written as
\begin{equation*}
 \kappa_2^{s_2} \me s_2\, c_1 \eq -\, \sum_{a=3}^{n}\, ( s_3 \cdots s_{a}) \, \kappa_a^{s_a}\,.
\end{equation*}
Following the same reasoning that we have just used, the left hand side of the latter equation depends just on $s_2$ while the terms on the right hand side clearly do not depend on $s_2$, we can thus conclude that
\begin{equation*}
 \kappa_2^{s_2}  \eq s_2 \,c_1 \me c_2 \quad \textrm{and} \quad  \sum_{a=3}^{n}\, ( s_3 \cdots s_{a}) \, \kappa_a^{s_a} \eq c_2 \,,
\end{equation*}
where  $c_2$ is a constant that does not depend on $\{s\}$. Following the same procedure until reaching the term $\kappa_n^{s_n}$, we end up with the following final result that solves the constraint (\ref{SumLambda2}):
\begin{equation}\label{Lambdaaca}
  \kappa_a^{s_a} \eq s_a\,c_{a-1} \me c_a \quad \textrm{with} \quad c_0 \eq c_n \eq 0 \,.
\end{equation}
Concerning the constants $c_1$, $c_2$, $\cdots$, $c_{n-1}$, they are arbitrary. Thus, in this problem we have $(n - 1)$ separation constants.  Finally, inserting Eqs. (\ref{ha}), (\ref{eta-lambda}) and (\ref{Lambdaaca}) into Eq. (\ref{Dif.EQ3}) we arrive at the following equations:
\begin{align}
      \slashed{D}_1^{s_1}  \hat{\psi}_{1}^{(-s_{1})} & \eq i\,(s_1\,\hat{m}_1 - c_1)  \hat{\psi}_{1}^{s_{1}} \nonumber \\
      \quad  \label{Dif.EQ4}\\
      \slashed{D}_a^{s_a}  \hat{\psi}_{a}^{(-s_{a})} & \eq i\,(s_a\,c_{a-1} - c_a)  \hat{\psi}_{a}^{s_{a}}  \;,\;\; \textrm{if} \quad a \geq 2 \,,\nonumber
\end{align}
where the operators $\slashed{D}_a^{s_a}$ were defined in Eq. (\ref{Dslash}). The above set of equations provides two equations for each $a$, one for $s_a=+1$ and the other for $s_a=-1$. Manipulating these two equations one easily obtain second order differential equations for $\hat{\psi}_{a}^{s_{a}}$, achieving the separability of the generalized Dirac equation that we are looking for. For appropriate boundary conditions, the constants $\{c_1,\cdots,c_{n-1}\}$ can only take discrete values. The general solution of Eq. (\ref{Dif.EQ1}) is, then, a linear combination of the particular solutions for each of the possible ``eigenvalues'' $\{c_1,\cdots,c_{n-1}\}$. In the next section, we shall use these results to separate the Dirac equation in some black hole spacetimes.

\section{Black Hole Spacetimes}\label{Sec.BHs}

In this section we shall separate the Dirac equation corresponding to a massive and electrically charged field of spin $1/2$ on the background of black holes whose horizons have topology $\mathbb{R}\times  S^2 \times \cdots \times S^2$. These black hole solutions, that possess electric and magnetic charge, have been obtained in Refs. \cite{Batista-BH,Maeda:2010qz,Ortaggio:2007hs} and are given by
\begin{widetext}
\begin{equation}\label{BHsol}
 ds^2 = -\, f(r)^2 \, dt^{2} + \frac{dr^2}{f(r)^2} + r^2\, \sum_{\ell=2}^{n} (d\theta_\ell^2 + \sin^2\theta_\ell\, d\phi_\ell^2)  \,,
\end{equation}
where  $f=f(r)$ is the following function of the coordinate $r$:
\begin{equation}\label{f(r)}
  f(r) = \sqrt{ \frac{1}{d-3} - \frac{2\,M}{r^{d-3}} + \frac{Q_e^2 (d-3)}{2(d-2)\,r^{2(d-3)}}-\frac{Q_m^2}{4(d-5)\,r^2} - \frac{\Lambda\,r^2}{d-1} } \,.
\end{equation}
\end{widetext}
In the latter expression, $d=2n$ is the dimension of the spacetime, $M$, $Q_e$ and $Q_m$ are the mass, the electric charge and the magnetic charge of the black hole respectively, while $\Lambda$ is the cosmological constant, see Ref. \cite{Batista-BH} for details. This spacetime is the solution of Einstein-Maxwell equations  with a cosmological constant $\Lambda$ and electromagnetic field $\bl{\mathcal{F}} =  d \bl{A}$, where the gauge field $\bl{A}$ is given by
\begin{equation*}
 \bl{A} = \frac{Q_e}{r^{d-3}}\,dt \ma Q_m \sum_{\ell=2}^{n} \cos\theta_\ell \, d\phi_\ell \,.
\end{equation*}
A suitable orthonormal frame for such a spacetime is given by
\begin{equation*}
  \left\{
    \begin{array}{ll}
      \bl{E}^1 \eq i\,f\,dt \;, \quad  \bl{E}^{\tilde{1}} \eq f^{-1}\,dr \,, \\
      \quad \\
      \bl{E}^\ell \eq r\,\sin\theta_\ell\, d\phi_\ell \;, \quad  \bl{E}^{\tilde{\ell}} \eq r\,d\theta_\ell \,,
    \end{array}
  \right.
\end{equation*}
where, as explained earlier, the index $\ell$ ranges from $2$ to $n$. Using this frame, the line element is given by
\begin{equation*}
  ds^2 \eq \sum_{a=1}^n \, ( \bl{E}^a \bl{E}^a \ma   \bl{E}^{\tilde{a}} \bl{E}^{\tilde{a}} ) \,.
\end{equation*}
In its turn, the gauge field can be written as
\begin{equation*}
  \bl{A} \eq A_a \, \bl{E}^a \ma A_{\tilde{a}} \, \bl{E}^{\tilde{a}}   \,,
\end{equation*}
where
\begin{equation}\label{A-BH}
  A_1 = \frac{-\,i\,Q_e}{f \, r^{d-3}} \;,\;\, A_\ell = \frac{Q_m}{r}\,\cot\theta_\ell \;,\;\,  A_{\tilde{a}} = 0 \,.
\end{equation}

A field of spin $1/2$ with electric charge $q$ and mass $m$ minimally coupled to the electromagnetic field and propagating in this spacetime obeys the following version of the Dirac equation:
\begin{equation}\label{DiracEq.BH}
  \gamma^\alpha\,(\nabla_\alpha - i\,q\,A_\alpha )\bl{\psi} \eq m \, \bl{\psi} \,.
\end{equation}
Using the definition of the Dirac operator, $D = \gamma^\alpha \nabla_\alpha$ along with the fact that $A_{\tilde{a}} = 0$, it follows that the above equation is written as
\begin{equation}\label{DiracEq.BH2}
  D\bl{\psi} \eq (\, m \ma i \,q  \,  A_a \gamma^a\,)\, \bl{\psi} \,.
\end{equation}
The aim of the present section is to integrate this equation.

In Sec. \ref{SecSeparability}, we have been able so separate an analogous equation for spaces that are the direct product of bidimensional spaces. However, the black hole line  element (\ref{BHsol}) is not of this special type, due to the warping factor $r^2$ in front of the angular part of the metric. Nevertheless, the conformal transformation
\begin{equation*}
  ds^2 \,\longrightarrow\, d\hat{s}^2 \eq \Omega^2\, ds^2 \quad \textrm{with} \;\quad \Omega = r^{-1}  
\end{equation*}
lead us to the following line element that is the direct product of bidimensional spaces:
\begin{equation}\label{BHsol-conf}
    d\hat{s}^2 = - \frac{f^2}{r^2} dt^{2} + \frac{dr^2}{(r f)^2} + \sum_{\ell=2}^{n} (d\theta_\ell^2 + \sin^2\theta_\ell d\phi_\ell^2) \,.
\end{equation}
A suitable orthonormal frame for this space is given by
\begin{equation}
  \left\{
    \begin{array}{ll}
      \hat{\bl{E}}^1 \eq i\,f\,r^{-1}\,dt \;, \quad  \hat{\bl{E}}^{\tilde{1}} \eq (r f)^{-1}\,dr \,,\\
      \quad  \label{FrameBH-conf}  \\
      \hat{\bl{E}}^\ell \eq \sin\theta_\ell\, d\phi_\ell \;, \quad  \hat{\bl{E}}^{\tilde{\ell}} \eq d\theta_\ell \,.
    \end{array}
  \right.
\end{equation}
The non-vanishing components of the spin connection are
\begin{equation}
  \left\{
    \begin{array}{ll}
       \hat{\omega}_{1\tilde{1}1} = - \, \hat{\omega}_{11\tilde{1}} = r\,f'- f   \,,\\
      \quad \label{SpinCon-BHconf} \\
      \hat{\omega}_{\ell\tilde{\ell}\ell} = - \, \hat{\omega}_{\ell\ell\tilde{\ell}} = \cot\theta_\ell \,,
    \end{array}
  \right.
\end{equation}
\\
where $f'$ stands for the derivative of $f$ with respect to its variable $r$. Using the conformal transformation of the Dirac operator, obtained in Sec. \ref{Sec.Conformal}, we can write the field equation (\ref{DiracEq.BH2}) in terms of an equation in the space with line element $ d\hat{s}^2$, so that the separability results of Sec. \ref{SecSeparability} can be fully used. Indeed, defining
\begin{equation}\label{psiNovo}
  \hat{\bl{\psi}} \eq \Omega^{( \frac{1}{2} - n )}\,  \bl{\psi} \eq  r^{( n - \frac{1}{2} )}\,  \bl{\psi} \,,
\end{equation}
and using  Eq. (\ref{DiracOp-conformal}), it follows that the field equation (\ref{DiracEq.BH2}) can be written as
\begin{equation}\label{DiracEq.BH-conf}
  \hat{D} \hat{\bl{\psi}} \eq \Omega^{-1}\,(\, m \ma i\, q \, A_a \gamma^a\,)\, \hat{\bl{\psi}}    \,.
\end{equation}
Then, defining
\begin{equation}\label{mA-BH}
  \hat{m} = \hat{m}_1 = r\,m \;\;, \quad \hat{A}_a = i\,q\,r\,A_a \;\;, \quad \hat{A}_{\tilde{a}} =  0  \,,
\end{equation}
it follows that Eq. (\ref{DiracEq.BH-conf}) takes exactly the form of the equation studied in Sec. \ref{SecSeparability}, namely we obtain Eq. (\ref{Dif.EQ1}). Moreover, and foremost, defining the coordinates
\begin{equation}\label{x-BH}
  x^1 = t \; ,\;\; y^{1} = r \; ,\;\; x^\ell = \phi_\ell \; ,\;\; y^{\ell} = \theta_\ell \,,
\end{equation}
it follows that the function $\hat{m}$ and the gauge field $\hat{A}_\alpha$ are exactly of the form necessary to attain separability, namely the constraints (\ref{Am}) and (\ref{ha}) are obeyed. Therefore, due to Eqs. (\ref{SpinBasis2}), (\ref{SpinorSeparable}) and (\ref{psiNovo}), it follows that a solution for Eq. (\ref{DiracEq.BH}) in the black hole background is provided by
\begin{widetext}
\begin{equation}\label{DiracSpinor}
 \bl{\psi} \eq r^{(\frac{1}{2} - n)}\, \sum_{\{s\}} \psi_1^{s_1}(t,r) \psi_2^{s_2}(\phi_2,\theta_2)\cdots \psi_n^{s_n}(\phi_n,\theta_n)
  \; \xi^{s_1}\otimes  \xi^{s_2}\otimes  \cdots \otimes  \xi^{s_n} \,.
\end{equation}
From Eqs. (\ref{Dslash}), (\ref{Dif.EQ4}), (\ref{A-BH}) and (\ref{FrameBH-conf})-(\ref{mA-BH}), it follows that the functions $\psi_{a}^{s_a}$ must be solutions of the following differential equations:
\begin{align}
      \left[ i\, s_1\,\left(  \frac{r}{i\,f} \partial_t -  \frac{q\,Q_e}{f \, r^{d-4}}  \right)
   +  \left( r\,f\,\partial_{r} + \frac{1}{2} (r\,f' - f)  \right) \right]  \psi_{1}^{(-s_{1})} & \eq i\,(s_1\,r\,m - c_1) \psi_{1}^{s_{1}} \nonumber \\
      \quad  \label{Dif.EQBH1}\\
     \left[ i \,s_\ell\, \left( \frac{1}{\sin\theta_\ell} \partial_{\phi_\ell} - i\,q\,Q_m\,\cot\theta_\ell \right)
   +  \left( \partial_{\theta_\ell} + \frac{1}{2} \cot\theta_\ell \right) \right]   \psi_{\ell}^{(-s_{\ell})} & \eq i\,(s_\ell\,c_{\ell-1} - c_\ell)  \psi_{\ell}^{s_{\ell}}   \,. \nonumber
\end{align}
As explained in Sec. \ref{SecSeparability}, the constants $c_1$, $c_2$, $\cdots$, $c_{n-1}$ are separation constants that generally take discrete values once boundary conditions and regularity requirements are imposed. The constant $c_n$, on the other hand, is zero.
Note that the coefficients in the above equations do not depend on the coordinates $t$ and $\phi_\ell$, which stems from the fact that these are cyclic coordinates of the metric, so that $\partial_t$ and $\partial_{\phi_\ell}$ are killing vector fields of both metrics $\bl{g}$ and $\hat{\bl{g}}$. Therefore, it is convenient to decompose the dependence of the fields $\psi_{a}^{s_a}$ on these coordinates in the Fourier basis, namely,
\begin{equation}\label{psiFourier}
   \psi_{1}^{s_1}(t,r) \eq e^{i \omega t} \, \Psi_1^{s_1}(r) \quad , \quad
 \psi_{\ell}^{s_\ell}(\phi_\ell, \theta_\ell) \eq e^{i \omega_\ell  \phi_\ell} \, \Psi_\ell^{s_\ell}(\theta_\ell) \,.
\end{equation}
The final general solution for the field $\bl{\psi}$ must, then, include a ``sum'' over all values of the Fourier frequencies $\omega$ and $\omega_\ell$ with arbitrary Fourier coefficients.  While $\omega$ can be interpreted as related to the energy of the field, $\omega_\ell$ are related to angular momentum. Note that in order to avoid conical singularities in the spacetime, the coordinates $\phi_\ell$ must have period $2\pi$, namely $\phi_\ell$ and $\phi_\ell + 2\pi$ should be identified. As it is well-known, a spin $1/2$ field changes its sign after a $2\pi$ rotation, which implies that the angular frequencies $\omega_\ell$ must be half-integers \cite{Abrikosov:2001nj,Camporesi:1995fb}:
\begin{equation}\label{AngularFrequency}
  \omega_\ell \eq \pm\, \frac{1}{2}\,,\,\,   \pm\, \frac{3}{2}\,,\,\, \pm\, \frac{5}{2}\,,\,\, \cdots  \;.
\end{equation}
Finally, inserting the decomposition (\ref{psiFourier}) into Eq. (\ref{Dif.EQBH1}), we end up with the following pairwise coupled system of differential equations:
\begin{align}
      \left[   r\,f\, \frac{d}{dr} + \frac{1}{2} (r\,f' - f)   +
i\,s_1\,\left(   \frac{\omega \,r}{f}  -  \frac{ q\,Q_e}{f \, r^{d-4}}  \right) \right]  \Psi_{1}^{(-s_{1})}
& \eq i\,(s_1\,m\,r -  c_1) \Psi_{1}^{s_{1}} \nonumber \\
      \quad  \label{Dif.EQBH2}\\
     \left[   \frac{d}{d\theta_\ell}   + \frac{1}{2} \cot\theta_\ell  -
s_\ell \left( \frac{ \omega_\ell}{\sin\theta_\ell}  - q\,Q_m\,\cot\theta_\ell \right) \right]   \Psi_{\ell}^{(-s_{\ell})}
& \eq i\,(s_\ell\,c_{\ell-1} - c_\ell)  \Psi_{\ell}^{s_{\ell}}   \,. \nonumber
\end{align}
\end{widetext}

\subsection{The angular part of Dirac's Equation}

Now, we shall investigate a little further the above equations. Let us start with the angular part of the equations, namely the equations
for $\Psi_{\ell}^{ s_{\ell}}$. One can make a simplification on these equations by performing a field redefinition along with a redefinition of the separation constants, as we show in the sequel. Instead of using the $n-1$ separation constants $c_1$, $c_2$, $\cdots$, $c_{n-1}$, we shall use the constants $\lambda_2$, $\lambda_3$, $\cdots$, $\lambda_{n}$, defined by
\begin{equation}\label{mu-c}
  \lambda_\ell \,\equiv\, \sqrt{c_{\ell-1}^2 \me c_{\ell}^2 } \,,
\end{equation}
where it is worth recalling that $c_n=0$, by definition. Inverting these relations, we find that the old constants can be written in terms of the new constants as follows:
\begin{equation}\label{c-mu}
 c_{\ell-1} \eq \sqrt{ \lambda_{\ell}^2 + \lambda_{\ell+1}^2 + \cdots + \lambda_{n}^2  } \,.
\end{equation}
Then, defining the parameter
\begin{equation*}
  \zeta_\ell \eq \arctanh(c_{\ell}/c_{\ell-1}) \,,
\end{equation*}
we find that
\begin{equation*}
  c_{\ell-1} \eq \lambda_{\ell} \cosh\zeta_\ell  \;\; \textrm{and} \;\;   c_{\ell} \eq \lambda_{\ell} \sinh\zeta_\ell\,,
\end{equation*}
so that the following relation holds:
\begin{equation*}
  s_\ell\,c_{\ell-1} \me c_\ell \eq s_\ell\,\lambda_{\ell}\,e^{-s_\ell \zeta_\ell}\,.
\end{equation*}
Thus, performing the field redefinition given by
\begin{equation}\label{FieldRedfAng}
  \Psi_{\ell}^{s_{\ell}}(\theta) \eq e^{s_\ell \zeta_\ell/2 }\, \Phi_{\ell}^{s_{\ell}}(\theta) \,,
\end{equation}
it turns out that the angular part of Eq. (\ref{Dif.EQBH2}) can be written in the following simpler way in terms of the fields $\Phi_{\ell}^{s_{\ell}}(\theta)$:
\begin{widetext}
\begin{equation}\label{Angular1}
   \left[     \frac{d}{d\theta_\ell}  + \frac{1}{2} \cot\theta_\ell  -
s_\ell \left( \frac{ \omega_\ell}{\sin\theta_\ell}  - q\,Q_m\,\cot\theta_\ell \right) \right]   \Phi_{\ell}^{(-s_{\ell})}
 \eq i\,s_\ell\,\lambda_{\ell} \Phi_{\ell}^{s_{\ell}} \,.
\end{equation}
\end{widetext}
Although it may seem that we did not achieve much simplification by the redefinition of the fields and separation constants, it turns out that in the case in which the black hole has vanishing magnetic charge, $Q_m = 0$, these equations reduce to
\begin{equation*}
   \left[     \frac{d}{d\theta_\ell}  + \frac{1}{2} \cot\theta_\ell  -
 \frac{s_\ell\, \omega_\ell}{\sin\theta_\ell} \right]   \Phi_{\ell}^{(-s_{\ell})}
 \eq i\,s_\ell\,\lambda_{\ell} \Phi_{\ell}^{s_{\ell}} \,.
\end{equation*}
In the latter form, the angular equations are identical to the equation $D_{S^2}\bl{\Phi} = i\,\lambda\,\bl{\Phi}$, where $D_{S^2}$ is the Dirac operator in the 2-dimensional unit sphere. To check this claim, one should use the frame $\bl{e}^1 \eq \sin\theta\, d\phi$ and   $\bl{e}^2 \eq d\theta$  along with the Dirac matrices $\gamma^1 = \sigma_1$ and $\gamma^2 = \sigma_2$. The solutions of the eigenvalue equation $D_{S^2}\bl{\Phi} = i\,\lambda\,\bl{\Phi}$ are well-known, the components of the 2-component spinor $\bl{\Phi}$ are written in terms of Jacobi polynomials \cite{Camporesi:1995fb,Abrikosov:2002jr}. From a geometrical point of view, these solutions can be understood in terms of the Wigner elements of the group $Spin(\mathbb{R}^3)$, that give rise to the so-called spin weighted spherical harmonics \cite{Boyle:2016tjj,Goldberg:1966uu}, which are tensorial generalizations of the spherical harmonics.    Moreover, the allowed eigenvalues $\lambda_\ell$ are also known, they must be non-zero integers \cite{Camporesi:1995fb,Trautman-Sphere}:
\begin{equation*}
  \lambda_\ell \eq \pm 1\,,\,\, \pm 2  \,,\,\, \pm 3 \,,\,\, \cdots \,\,.
\end{equation*}
Solutions with non-integer eigenvalues are not well-defined on the whole sphere, while a vanishing eigenvalue is forbidden by the Lichnerowicz theorem \cite{Lichnerowicz}, since the sphere is a compact manifold with positive curvature.

Regarding the general case in which the black hole magnetic charge is non-vanishing,  $Q_m \neq 0$, we have tried to make a redefinition of the fields $\Phi_{\ell}^{s_\ell}$ by means of a general linear combination of the fields $\Phi_{\ell}^{+}$ and $\Phi_{\ell}^{-}$, with non-constant coefficients, in order to convert Eq. (\ref{Angular1}) into the eigenvalue equation $D_{S^2}\bl{\Phi} = i\,\lambda\,\bl{\Phi}$. However, it turns out that the coefficients of the linear combination must obey fourth-order differential equations, whose solutions seem to be quite difficult to attain analytically. In spite of this, we can make an important progress regarding the system of equations (\ref{Angular1}) by decoupling the fields $\Phi_{\ell}^{+}$ and $\Phi_{\ell}^{-}$, which, after all, is our goal at this paper. The final results is that the fields $\Phi_{\ell}^{s_\ell}$ satisfy the following second order differential equation:
\begin{widetext}
\begin{equation}\label{SecondOrder-Ang}
  \frac{1}{\sin\theta_\ell}\frac{d}{d\theta_\ell}\left( \sin\theta_\ell \frac{d \Phi_{\ell}^{s_\ell} }{d\theta_\ell} \right) \ma
  \left[\,  \frac{ (1+2 q Q_m)\omega_\ell\,\cos\theta_\ell}{\sin^2\theta_\ell} -
    \frac{1+2 q Q_m + 2 \omega_\ell^2}{2\,\sin^2\theta_\ell }  +
    \frac{(1- 4 q^2 Q_m^2) \,\cos^2\theta_\ell }{4\sin^2\theta_\ell} -
     \lambda_\ell^2\,\right]\Phi_{\ell}^{s_\ell} \eq 0\,.
\end{equation}
\end{widetext}
It is worth stressing that the latter equation must be supplemented by the requirement of regularity of the fields $\Phi_{\ell}^{s_\ell}$ at the points $\theta_\ell=0$ and $\theta_\ell=\pi$, where our coordinate system breaks down. These regularity conditions transform the task of solving the latter equation in a Sturm-Liouville problem, so that the possible values assumed by the separation constants $\lambda_\ell$ form a discrete set. Since the case $Q_m = 0$ in Eq. (\ref{SecondOrder-Ang}) has a known solution, as described above, it follows that we can look for solutions for the case $Q_m\neq 0$ by means of perturbation methods, with $Q_m$ being the perturbation parameter. Indeed, in the celebrated paper \cite{Press:1973zz}, a similar path has been taken by Press and Teukolsky in order find the solutions and their eigenvalues for the angular part of the equations of motion for fields with arbitrary spin on Kerr spacetime, in which case the angular momentum of the black hole was the order parameter. In this respect, see also Ref. \cite{Chakrabarti}.

\subsection{The radial part of Dirac's Equation}

In order to solve the pair of radial equations in (\ref{Dif.EQBH2}), we should first decouple the fields  $\Psi_{1}^{+}$ and  $\Psi_{1}^{-}$. This can be easily attained by defining
\begin{align*}
   B_{s_1}(r) &\eq \frac{1}{r\,f}\left[ \frac{1}{2}(r\,f' - f) - i s_1 \left(   \frac{\omega \,r}{f}  -  \frac{ q\,Q_e}{f \, r^{d-4}}  \right)  \right]  \,,\\
  C_{s_1}(r)  &\eq -\,\frac{i}{r\,f} (s_1\, m \, r + c_1) \,,
\end{align*}
in terms of which the radial equation in (\ref{Dif.EQBH2}) can be written as
\begin{equation}\label{Dirac-Radial}
  \frac{d}{dr} \Psi_{1}^{s_{1}} \eq   -\,   B_{s_1}\,\Psi_{1}^{s_{1}} \ma  C_{s_1}\,\Psi_{1}^{-s_{1}}  \,.
\end{equation}
Then, deriving this equation with respect to $r$ and using Eq. (\ref{Dirac-Radial}) to substitute $\Psi_{1}^{-s_{1}} $ in terms of $\Psi_{1}^{s_{1}}$, we are eventually led to the following decoupled second order differential equation:
\begin{widetext}
\begin{equation}\label{Radial_2ndOrder}
   \frac{d^2 \Psi_{1}^{s_{1}}}{dr^2} \ma
   \left( B_{s_1} + B_{-s_1} - \frac{1}{C_{s_1}} \frac{dC_{s_1}}{dr} \right) \left(\frac{d \Psi_{1}^{s_{1}}}{dr} + B_{s_1} \Psi_{1}^{s_{1}}  \right) \ma
   \left( \frac{dB_{s_1}}{dr}  - B_{s_1}^2  - C_{s_1} C_{-s_1}\right)\Psi_{1}^{s_{1}} \eq 0 \,.
\end{equation}
\end{widetext}

An analytical exact solution of the latter differential equation is, probably, out of reach. Nevertheless, we can use Eq. (\ref{Radial_2ndOrder}) to infer the asymptotic forms of the solution near the infinity, $r \rightarrow \infty$, as well as near the horizon $r \rightarrow r_\star$, where $r_\star$ is a root of the function $f$, namely $f(r_\star)=0$. Particularly, we shall prove in the sequel that, in the case of vanishing cosmological constant, the well-known case $d=4$ is qualitatively different from the higher-dimensional cases $d\geq 6$. In order to do this analysis, we shall write Eq. (\ref{Radial_2ndOrder}) as
\begin{equation}\label{h0h1}
  \frac{d^2 \Psi_{1}^{s_{1}}}{dr^2} \ma h_{1}^{s_1}(r) \,\frac{d \Psi_{1}^{s_{1}}}{dr} \ma   h_0^{s_1}(r)\,\Psi_{1}^{s_{1}} \eq 0 \,,
\end{equation}
where $h_1^{s_1}$ and $h_0^{s_1}$ are defined by comparing Eqs. (\ref{Radial_2ndOrder}) and (\ref{h0h1}). Then, we can work out the asymptotic forms of the coefficients $h_1^{s_1}$ and $h_0^{s_1}$ in the region of interest. In particular, for a consistent investigation of the asymptotic form of the solutions of Eq. (\ref{h0h1}) in the limit $r\rightarrow\infty$, if we want to know $\Psi_{1}^{s_{1}}$ up to order $r^{-p}$ we need to know $h_1^{s_1}$ up to order $r^{-(p+1)}$ and consider $h_0^{s_1}$ up to order $r^{-(p+2)}$.

Looking at the function $f(r)$ in Eq. (\ref{f(r)}), we see that the term that multiplies $\Lambda$ becomes the dominant one as we approach the infinity, $r \rightarrow \infty$. Therefore, it is intuitive to guess that the cases of vanishing and non-vanishing $\Lambda$ should be qualitatively different. Thus, let us separate the analysis of these two cases.

First, let us consider the case $\Lambda \neq 0$. Collecting the coefficients that multiply $\frac{d \Psi_{1}^{s_{1}}}{dr}$ and $\Psi_{1}^{s_{1}}$ in Eq. (\ref{Radial_2ndOrder}) and then expanding them in powers of $r^{-1}$, we can find, after some algebra, the following asymptotic forms:
\begin{align*}
 h_0^{s_1}(r) & =  \frac{(d-1) m^2}{\Lambda\, r^2}   +  \frac{i\,(d-1)\,s_1\,\omega}{\Lambda \, r^3} + O\left( \frac{1}{r^4} \right)  , \\
 \quad \\
 h_1^{s_1}(r) & = \frac{1}{r} - \frac{s_1\,c_1}{m\, r^2}  +
  \left[ \frac{2(d-1)}{(d-3)\Lambda} -  \frac{c_1^2}{m^2}\right] \frac{1}{r^3} + O\left( \frac{1}{r^4} \right) .
\end{align*}
In particular, considering the expansion of $h_0^{s_1}$ up to order $r^{-2}$ and the expansion of $h_1^{s_1}$ up to order $r^{-1}$, we are led to the following asymptotic form:
\begin{equation*}
  \Psi_{1}^{s_{1}}(r) \, \sim \, C_0\,\sin\left[ \frac{m \sqrt{d-1}}{\sqrt{\Lambda}} \,\log(r)  \ma \varphi_0 \right] + O\left( \frac{1}{r} \right)\,,
\end{equation*}
where $C_0$ and $\varphi_0$ are arbitrary integration constants. Note, however, that the field that is ``the solution'' of the Dirac equation, $\bl{\psi}$, has a further decaying multiplicative factor $r^{(1-d)/2}$, in accordance with Eq. (\ref{DiracSpinor}).

Now, let us consider the context of vanishing cosmological constant, $\Lambda = 0$. In such a case, one can see, after some algebra, that the asymptotic forms of the functions $h_0$ and $h_1$ are the following when $d\geq 6$:
\begin{align*}
 h_0^{s_1}(r) & =  \left[ (d-3)^2\,\omega^2 - (d-3)\,m^2  \right] \,  +     \\
   &    \left[ \frac{3}{4} + (d-3)\left(c_1^2 -  \frac{i \omega c_1 }{m}\right) \right. \, - \\
 &    \frac{(d-3)^2 Q_m^2 m^2}{4(d-5)} +  \left. \frac{(d-3)^3 Q_m^2 \omega^2}{2(d-5)} \right]\frac{1}{r^2}  +   O\left( \frac{1}{r^3} \right)\,,  \\
 \quad \\
 h_1^{s_1}(r) & = -\,\frac{1}{r} - \frac{s_1\,c_1}{m\, r^2}  +   O\left( \frac{1}{r^3} \right) .
\end{align*}
However, these formulas do not apply to the well-studied situation $d=4$, in which case $h_0^{s_1}$ has a term of order $r^{-1}$ depending on $M$ and $Q_e$ and the coefficient of order $r^{-2}$ has additional terms also depending on $M$ and $Q_e$ . Analogously, for $d=4$, the function $h_1^{s_1}$ also has additional contributions of order $r^{-2}$ stemming from the mass and the electric charge of the black hole. Thus, we conclude that, for $\Lambda = 0$, the spinor field that represents a charged particle of spin $1/2$ moving in the black hole (\ref{BHsol}) has qualitatively different fall off properties in the asymptotic infinity depending on whether $d=4$ or $d\geq 6$.

A similar asymptotic analysis can be performed near the horizons, namely near the values of $r$ for which the function $f$ vanishes. In such a case the coordinate $r$ ceases to be reliable and we should change the radial coordinate to tortoise-like coordinates, see \cite{Press:1973zz} for instance. Besides such asymptotic behaviours, one can also look for approximate solutions valid in a broader domain by means of other approximation methods. For instance, in Ref.\cite{Mukhopadhyay:2000cp} an approximate solution for the Dirac field on the Kerr spacetime has been obtained using the WKB method after transforming the radial second order differential equation into a Schr\"{o}dinger equation, see also \cite{Schutz:1985zz,Berti:2014bla}.


\section{Conclusions and Perspectives}\label{Sec.Conclusions}

In this article we have shown that the Dirac Equation coupled to a gauge field can be decoupled in even-dimensional manifolds that are the direct product of bidimensional spaces, provided that the gauge field is also ``separated'' in accordance with the bidimensional blocks, as shown in Eq. (\ref{Am}). Then, we have used this fact along with the conformal transformation of the Dirac operator to decouple the equation of motion of a charged test field of spin $1/2$ propagating in the background of the black hole solution (\ref{BHsol}). We have shown that the latter problem reduces to solving a second order radial differential equation, whose asymptotic behaviour has been worked out, along with a second order angular differential equation with regularity conditions. In particular, we have argued that if the black hole has vanishing magnetic charge then the angular equation reduces to the eigenvalue problem for the Dirac operator on the sphere, whose solutions are known.

The separation attained in the present work paves the way to analyse the quasi-normal modes associated to a field of spin $1/2$ on the background of the black hole considered here. Moreover, in four dimensions, considering the case of positive cosmological constant and taking the limit of equal temperatures for the black hole horizon and the cosmological horizon we end up with the so-called Nariai spacetime \cite{Batista-BH}, whose geometric structure is much simpler than the black hole solution. Therefore, hopefully, one can use the tools presented here to find, analytically, the quasi-normal modes of a spin $1/2$ field on higher-dimensional versions of the Nariai spacetime. This line of research is of physical relevance both from the theoretical and experimental points of view. Indeed, quasi-normal modes are related to the analysis of stability of black holes and their knowledge are of relevance on applications of the AdS/CFT correspondence. Furthermore, these modes play a central role on the measurements of gravitational radiation as well as on the characterization of astronomical objects \cite{Kokkotas:1999bd,Berti:2009kk, Nollert:1999ji}. Therefore, we intend to continue our work addressing these points in the near future. In addition, we aim to investigate the possibility of superradiance phenomenon in these spacetimes in forthcoming works.

\begin{acknowledgments}
J. V. thanks CNPq for the financial support. 
\end{acknowledgments}


\end{document}